\documentclass[aps,prd,reprint,superscriptaddress,amsmath,amssymb,nofootinbib,floatfix]{revtex4-2}

\usepackage{graphicx}
\usepackage{dcolumn}
\usepackage[utf8]{inputenc}
\usepackage{xcolor}
\usepackage[unicode=true,colorlinks,linkcolor=blue,anchorcolor=blue,urlcolor=blue,citecolor=blue,breaklinks=true]{hyperref}

\newcommand{\pararrow}{\mathord{\buildrel{\lower3pt\hbox{$\scriptscriptstyle\leftrightarrow$}}\over {\partial}}} 
\newcommand{\pararrowk}[1]{\mathord{\buildrel{\lower3pt\hbox{$\scriptscriptstyle\leftrightarrow$}}\over {\partial}\hspace*{-0.18em}{}^#1}\hspace*{-0.18em} \,} 

\usepackage{physics}
\usepackage{bm}

\newcommand{\qfnu}{\affiliation{College of Physics and Engineering, Qufu Normal University, Qufu 273165, China}}

\newcommand{\hnnu}{\affiliation{Institute of Particle and Nuclear Physics, Henan Normal University, Xinxiang 453007, China}}

\begin{document}
	
\title{Radiative decays of $X(3872)$ in $D{\bar D}^*$ molecule scenario}
	
	\author{Fan Wang} \qfnu
	\author{Gang Li}\email{gli@qfnu.edu.cn} \qfnu
	\author{Shi-Dong Liu} \email{liusd@qfnu.edu.cn}\qfnu
	\author{Qi Wu}\email{wuqi@htu.edu.cn} \hnnu
	
\begin{abstract}
We investigate the radiative decays of the $X(3872)$ to $\gamma V~(V=\rho^0,\, \omega)$ in the molecule scenario, where the $X(3872)$ is regarded as a pure hadronic molecule of the $D\bar{D}^*+c.c$ in an $S$-wave with the quantum numbers $J^{PC}=1^{++}$. The radiative processes were assumed to occur via the triangle hadronic loops, and the relevant calculations were conducted using an effective Lagrangian approach. It is found that the absolute decay widths are model-dependent, but the relative width ratio is rather independent of the model parameter. Moreover, the calculated results indicate that the radiative decays of the $X(3872)$ are strongly influenced by the molecular configuration characterized by the proportion of the charged and neutral constituents. We hope that the present calculations could be tested by the experimental measurements.		
\end{abstract}

\date{\today}

\maketitle

\section{Introduction}\label{sec:Introduction}

Since the Belle collaboration discovered $X(3872)$ in $\pi^+ \pi^- J/\psi$ invariant mass spectrum of $B^+\to K^+\pi^+ \pi^- J/\psi$ process in 2003~\cite{Belle:2003nnu}, many exotic state candidates have been discovered by various experimental groups around the world and have become important research subjects in the field of hadron physics (see some reviews~\cite{Chen:2016qju,Lebed:2016hpi,Esposito:2016noz,Dong:2017gaw,Guo:2017jvc,Olsen:2017bmm,Karliner:2017qhf,Brambilla:2019esw,Guo:2019twa,Yang:2020atz,Meng:2022ozq,Liu:2024uxn}).

The $X(3872)$, as the first observed charmonium-like state and one of the well-studied representatives of exotic states, has been thoroughly studied both experimentally and theoretically. It provides an opportunity to explore new physical phenomena, especially in the field of interactions between quark matter and hadrons~\cite{Zhang:2020dwn}.
To date, many explanations have been proposed to understand the particular properties of the $X(3872)$. 

It is worth noting that the final states in all the observed decay modes of the $X(3872)$ contain charm and anti-charm quarks. Therefore, it is believed that the quark components of $X(3872)$ include at least $c\bar{c}$, which means that the $X(3872)$ could be a good charmonium candidate. Given the quantum numbers $J^{PC}$ (its quantum numbers have been determined to be $J^{PC}=1^{++}$ \cite{LHCb:2013kgk,ParticleDataGroup:2024cfk}) and the mass of $X(3872)$ (the world average mass is 3871.65 MeV)~\cite{ParticleDataGroup:2024cfk}, the only choice is the $\chi_{c1}(2P)$ state~\cite{Barnes:2003vb,Eichten:2004uh,Chen:2007vu,Meng:2007cx,Liu:2007uj,Wang:2010ej}. However, the charmonium scenario met some difficulty when considering the resonance parameters and decay behaviors. Specifically, the $X(3872)$ mass is lower than the prediction of the Goldfrey–Isgur quark model by almost 80 MeV~\cite{Godfrey:1985xj} and the experimental ratio $\mathcal{B}[X\to \pi^+ \pi^- \pi^0 J/\psi]/\mathcal{B}[X\to \pi^+ \pi^- J/\psi]$ shows large isospin-breaking effects~\cite{BaBar:2010wfc,BESIII:2019qvy,Belle:2005lfc}. The former fact indicates that the $X(3872)$ is not a pure charmonium state, while the latter phenomenon implies that it contains a significant $c\bar{c}u\bar{u}$ component and appears a mixture of isoscalar and isovector components.

The properties of the $X(3872)$ cannot be simply explained in the context of conventional constituent quark models. The $X(3872)$ exhibits two notable features: a rather narrow width ($\Gamma < 1.2~\mathrm{MeV}$) and  a mass nearly coinciding with the $D^0 \bar{D}^{*0}$ threshold.
Analogous to the deuteron, a loosely bound state composed of a proton and a neutron, it is natural to regard the $X(3872)$ as a $D\bar{D}^*$ loosely bound state~\cite{Swanson:2003tb,Voloshin:2003nt,AlFiky:2005jd,Liu:2008fh,Sun:2011uh,Nieves:2012tt,Guo:2013sya,Karliner:2015ina, Braaten:2005ai,Gamermann:2009fv,Ortega:2009hj,Hanhart:2011tn,Li:2012cs,Takeuchi:2014rsa,Zhou:2017txt,Mutuk:2018zxs}.   The molecule picture can explain not only the coincidence of the mass of the $X(3872)$ with the $D^0 \bar{D}^{*0}$ threshold naturally, but also the isospin-breaking effects. Another multiquark configuration of the $X(3872)$ is the tetraquark interpretation with constituent $c\bar{c}q\bar{q}$~\cite{Maiani:2005pe,Ebert:2005nc,Cui:2006mp,Matheus:2006xi,Dubnicka:2010kz,Wang:2013vex}. Besides, a hybrid with constituent $c\bar{c}g$ has also been proposed to understand the inner structure of the $X(3872)$~\cite{Close:2003mb,Li:2004sta,Petrov:2005tp}. However, the threshold effect could be ruled out, since the $X(3872)$ has been observed in more than one decay channel.

To date, the interpretation of the nature of $X(3872)$ still hangs in doubt. The investigations of the mass spectrum as well as production or decay behaviors are both crucial for understanding the nature of the $X(3872)$. Studying the decay patterns and decay widths will shed light on the underlying structure of $X(3872)$ since they would depend on the structure assumption such as a charmonium~\cite{Meng:2007cx,Barnes:2005pb}, a hadronic molecule~\cite{Guo:2014taa,Swanson:2003tb,Ferretti:2014xqa,Wu:2021udi,Wang:2022qxe}, a tetraquark configuration~\cite{Wang:2023sii}, or even a mixed state of these components~\cite{Dong:2008gb,Dong:2009uf,Dong:2009yp}. Refs.~\cite{Dubynskiy:2007tj,Fleming:2008yn,Mehen:2015efa} propose to measure the pion transition of $X(3872)$ to $\chi_{cJ}$ to distinguish various interpretations. Radiative decays of the $X(3872)$ require the annihilation of the quark-antiquark pairs in the different configuration of the wave function, thus providing a unique way to distinguish the charmonium and molecular assignments for this state. In Refs.~\cite{Swanson:2003tb,Ferretti:2014xqa,Barnes:2005pb,Badalian:2012jz,Dong:2009uf}, the ratio $\mathcal{B}[X\to \psi^\prime \gamma]/\mathcal{B}[X\to J/\psi \gamma]$ is calculated under various assignments and the results are highly dependent on the structure of the $X(3872)$. In addition, it is found that the decays $X(3872)\to J/\psi\pi \gamma$
and $X(3872)\to J/\psi\pi \pi \gamma$ can be used to probe the
isoscalar and isovector components of the $X(3872)$ wave function~\cite{Wu:2022wru}.

Different from the radiative decays of the $Z^0_c(3900)$~\cite{Chen:2015igx} and $X(3872)\to \psi^{(\prime)} \gamma$, where light quark pairs annihilate into a photon and the
rest charm and anticharm quarks form a charmonium in the final states, the $X(3872)$ radiative decays to light hadrons may occur via charm and anticharm quark annihilation and the
rest light quark hadronization. Radiative decays to light hadrons may provide a surprising amount of insight into the unusual properties of the  $X(3872)$, especially in distinguishing the charmonium and molecular assignments.

The intermediate meson loop, a well-established nonperturbative transition mechanism~\cite{Li:1996yn,Cheng:2004ru,Guo:2010ak}, has been widely used to study the production and decays of exotic states \cite{Liu:2024ogo,Wang:2022aiu,Wu:2023rrp,Cai:2024glz}. In this work, we shall systematically investigate the radiative decay processes of $X(3872) \rightarrow \gamma V (V=\rho^0 , \omega)$ via the charmed meson loops using the effective Lagrangian approach, where the $X(3872)$ is assumed to be a pure mesonic molecule of the $\bar D D^\ast$ pair.

The rest of this paper is organized as follows. In Sec. \ref{sec:theoretical framework}, we present the theoretical framework used in this work. Then in Sec.\ref{sec:numerical results}, the numerical results are presented, and a brief summary is given in Sec. \ref{sec:summary}.

\section{Theoretical framework}\label{sec:theoretical framework}

\begin{figure*}
    \centering    \includegraphics[width=0.96\linewidth]{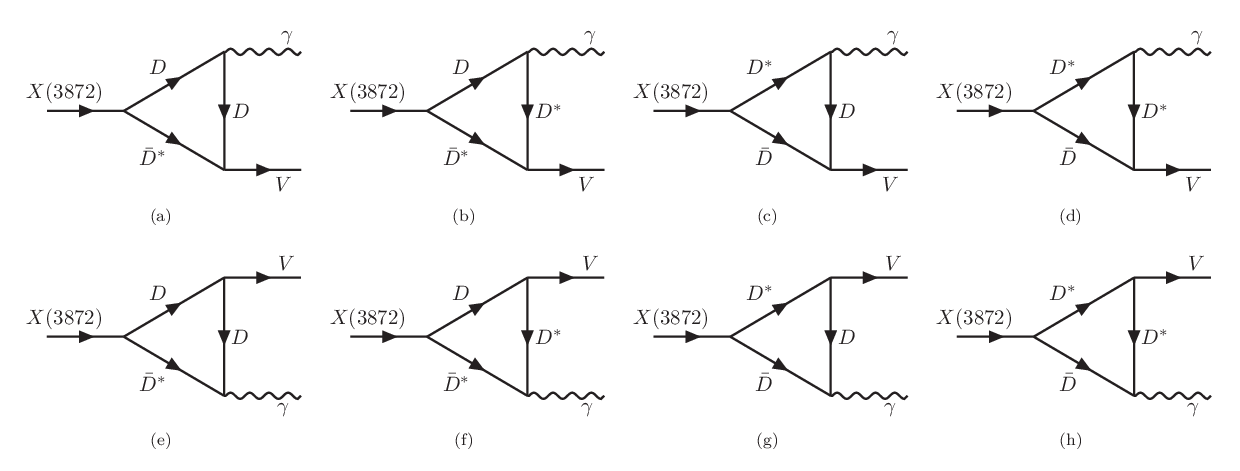}
    \caption{Feynman diagrams for the processes $X(3872)\rightarrow \gamma V(V=\rho^0,\, \omega)$ via charmed meson loops. The charge conjugated loops are not shown here, but have been taken into account in the calculations.}
    \label{fig:feyndiags}
\end{figure*}

Fig.~\ref{fig:feyndiags} shows the radiative decay processes of $X(3872)$ in $D{\bar D}^*$ molecule scenario via charmed meson loops. Firstly, the $X(3872)$ decays into a pair of charmed and anticharmed mesons. Then, the produced meson pair are rescattered into a photon and a vector meson by exchanging a proper charmed meson. In the following, we present the relevant effective Lagrangian and decay amplitudes for calculating the radiative decay processes of $X(3872)$.

\subsection{Effective interaction Lagrangians}\label{subsec:Effective interaction Lagrangians}

Here we assume that the $X(3872)$ is a pure hadronic molecule state of the $D\bar{D}^*+c.c$ in $S$-wave with the quantum numbers $J^{PC}=1^{++}$ \cite{Belle:2015qeg,Braaten:2007dw,Chen:2008cg,Meng:2014ota,Dunwoodie:2007be,Mehen:2011ds,Braaten:2020iye}. The wave function of the $X(3872)$ could be written as \cite{Achasov:2024ezv}
    \begin{align}\label{eq:WFX(3872)}
        |X(3872)\rangle&=\frac{\cos{\theta}}{\sqrt{2}}|D^{\ast 0}\bar D^{0}+D^{0}\bar D^{\ast 0}\rangle \nonumber\\
        &+\frac{\sin{\theta}}{\sqrt{2}}|D^{\ast +}D^{-}+D^{+}D^{\ast -}\rangle,
    \end{align}
where $\theta$ is a phase angle describing the proportion of neutral and charged constituents. The angle $\theta=0$ and $\theta=\pi/2$ denote the $X(3872)$ as a pure $D^{0}\bar D^{\ast 0}+c.c$ and $D^{+}D^{\ast -}+c.c$, respectively, while $\theta=\pi/4$ stands for the isospin singlet state \cite{Wang:2022qxe,Zheng:2024eia,Liu:2023gtx,Wu:2021udi}.

The interaction of the $X(3872)$ with a pair of charmed and anticharmed mesons $D^{\ast}\bar D$ is described by \cite{Wang:2022qxe}
\begin{align}\label{eq:LX(3872)}
    \mathcal{L}_X&=\frac{g_n}{\sqrt{2}}X_{\mu}^{\dagger}(D^{\ast 0\mu}\bar D^{0}+D^{0}\bar D^{\ast0\mu})\nonumber\\
    &+\frac{g_c}{\sqrt{2}}X_{\mu}^{\dagger}(D^{\ast+\mu}D^{-}+D^{+}D^{\ast-\mu}),
\end{align}
where $g_n$ and $g_c$ are the coupling constants of the $X(3872)$ with its neutral and charged components, respectively.

The couplings between the light vector and charmed mesons, obtained based on the heavy quark limit and chiral symmetry, are \cite{Wu:2021udi}
    \begin{align}\label{eq:Lagrangian for light hadron}
        \mathcal{L} &=-ig_{DDV}D_{i}^{\dagger}\pararrowk{\mu}D^{j}(V_{\mu})_{j}^{i}\nonumber\\
    &-2f_{D^{\ast}DV}\epsilon_{\mu\nu\alpha\beta}(\partial^{\mu}V^{\nu})_{j}^{i}(D_{i}^{\dagger}\pararrowk{\alpha}D^{\ast\beta j}-D_{i}^{\ast\beta\dagger} \pararrowk{\alpha} D^{j})\nonumber\\  &+ig_{D^{\ast}D^{\ast}V}D_{i}^{\ast\nu\dagger}\pararrowk\mu D_{\nu}^{\ast j}(V_{\mu})_{j}^{i}\nonumber\\
    &+4if_{D^{\ast}D^{\ast}V}D_{i\mu}^{\ast\dagger}(\partial^{\mu}V^{\nu}-\partial^{\nu}V^{\mu})_{j}^{i}D_{\nu}^{\ast j}+\mathrm{H.c.},
    \end{align}    
where the $ D^{(\ast)\dagger}=(\bar{D}^{(\ast) 0},D^{(\ast)-},D_{s}^{(\ast)-})$ is the charmed meson triplet. Since there are no strange mesons in the wave function of $X(3872)$, we do not need to consider the strange charmed meson loop in this work. The $V$ denotes the nonet vector mesons in the matrix form \cite{Wang:2022qxe,Zheng:2024eia,Liu:2023gtx,Wu:2021udi}
\begin{equation}\label{eq:vector matrix}
    \begin{aligned}
    V=\left(\begin{array}{ccc}
     \frac{\rho^0}{\sqrt{2}}+\frac{\omega}{\sqrt{2}}&\rho^+&K^{\ast +}\\
     \rho^-&-\frac{\rho^0}{\sqrt{2}}+\frac{\omega}{\sqrt{2}}&K^{\ast 0}\\
     K^{\ast -}&\bar{K}^{\ast 0}&\phi
\end{array}\right).
    \end{aligned}
\end{equation}

In the heavy quark and chiral limits, the coupling constants $g_{D^{(\ast)}D^{(\ast)}V}$ and $f_{D^{(\ast)}D^{(\ast)}V}$ are determined by the following relationship~\cite{Wang:2022qxe,Zheng:2024eia,Liu:2023gtx,Wu:2021udi},
\begin{align}\label{eq:couplings}
        g_{DDV}=g_{D^\ast D^\ast V}=\frac{\beta g_V}{\sqrt{2}}, \nonumber\\
        f_{D^\ast DV}=\frac{f_{D^\ast D^\ast V}}{m_{D^\ast}}=\frac{\lambda g_V}{\sqrt{2}}.
\end{align}

The interaction between the charmed mesons and the photon has two kinds of couplings,  corresponding to the electric and magnetic interactions~\cite{Shi:2023mer}, which are
    \begin{align}\label{eq:electric interactions}
        \mathcal{L}_{e}&=\partial_{\mu}D^{\dagger}\partial^{\mu}D-m_{D}^{2}D^{\dagger}D+ieQ_{D}A_{\mu}(\partial^{\mu}D^{\dagger}D-D^{\dagger}\partial^{\mu}D)\nonumber\\
        &+e^{2}Q_{D}^{2}A_{\mu}A^{\mu}D^{\dagger}D-\frac{1}{2}D_{\mu\nu}^{\ast\dagger}D^{\ast\mu\nu}+m_{D^{\ast}}^{2}D_{\mu}^{\ast\dagger}D^{\ast\mu}\nonumber\\
        &+ieQ_{D^{\ast}}A_{\mu}(D_{\nu}^{\ast\dagger}\pararrowk\mu D^{\ast\nu}+\partial_{\nu}D^{\ast\dagger\mu}D^{\ast\nu}-D^{\ast\dagger\nu}\partial_{\nu}D^{\ast\mu})\nonumber\\
        &-e^{2}Q_{D^{\ast}}^{2}(A_{\mu}A^{\mu}D_{\nu}^{\ast\dagger}D^{\ast\nu}-A_{\mu}A_{\nu}D^{\ast\dagger\nu}D^{\ast\mu}) \, ,
    \end{align}
and 
    \begin{align}\label{eq:magnetic interactions}
        \mathcal{L}_{m}&=-ieF^{\mu\nu}\left(D_{\mu}^{\ast\dagger}D_{\nu}^{\ast}-D_{\nu}^{\ast\dagger}D_{\mu}^{\ast}\right) \left(\frac{Q\beta^{\prime}}{2}-\frac{Q^{\prime}}{2m_c}\right)\nonumber\\        
        &+e\epsilon^{\mu\nu\alpha\beta}F_{\mu\nu}v_{\alpha} \left( D^{\dagger}D_{\beta}^{\ast}+D_{\beta}^{\ast\dagger}D \right) \left(\frac{Q\beta^{\prime}}{2}+\frac{Q^{\prime}}{2m_c}\right)\nonumber\\
        &+ieF^{\mu\nu}\left(\bar{D}_{\mu}^{\ast\dagger}\bar{D}_{\nu}^{\ast}-\bar{D}_{\nu}^{\ast\dagger}\bar{D}_{\mu}^{\ast} \right) \left(\frac{Q\beta^{\prime}}{2}-\frac{Q^{\prime}}{2m_c}\right)\nonumber\\        &+e\epsilon^{\mu\nu\alpha\beta}F_{\mu\nu}v_{\alpha}\left(\bar{D}^{\dagger}\bar{D}_{\beta}^{\ast}+\bar{D}_{\beta}^{\ast\dagger}\bar{D}\right) \left(\frac{Q\beta^{\prime}}{2}+\frac{Q^{\prime}}{2m_c}\right).
    \end{align}
    
Here $eQ_{D^{\ast}}$ represents the electric charge of the $D^{\ast}$ and $D_{\mu\nu}^{\ast}=\partial_{\mu}D_{\nu}^{\ast}-\partial_{\nu}D_{\mu}^{\ast}$. The light quark charge matrix $Q=\mathrm{diag}(2/3,-1/3)$, the charge of the charm quark $Q^{\prime}=2/3$, and $F_{\mu\nu}=\partial_\mu A_\nu-\partial_\nu A_\mu$ stands for the electromagnetic field tensor. $m_c=1863~\mathrm{MeV}$ is the charm quark mass, $v =(1,\textbf{0})$ is the four-velocity of the charmed meson \cite{Shi:2023mer,Hu:2005gf}. In addition, $\beta^{\prime -1}=379~\mathrm{MeV}$ \cite{Shi:2023mer,Hu:2005gf}.

\subsection{Amplitudes of $X(3872) \rightarrow \gamma V$}\label{Amplitudes of X to gamma V}

With the Lagrangian above, we can obtain the decay amplitude of the $X(3872) \rightarrow \gamma V$
    \begin{align}
        \mathcal{M}_{X}=\frac{1}{2}g_{n/c}
        \varepsilon_{\mu}(X)\varepsilon_{\beta}^{\ast}(V)\varepsilon_{\nu}^{\ast}(\gamma)I^{\mu\beta\nu},
    \end{align}
where the loop integral part for each diagram in Fig.~\ref{fig:feyndiags} can be written as 
    \begin{align}
        I_{(a)}^{\mu\beta\nu}&=\int\frac{d^4q}{(2\pi)^4}g^{\mu\sigma}[-eQ_D(q+p_1)^{\nu}]\nonumber\\
        &\times[2f_{D^\ast DV}\epsilon^{\alpha\beta\delta\eta}p_{4\alpha}(q-p_2)_{\delta}]S(p_1, m_D)\nonumber\\
        &\times S_{\sigma \eta}(p_2, m_{D^\ast})S(q,m_D)F(q^2,m_{D}^{2}),
    \end{align}
    \begin{align}
        I_{(b)}^{\mu\beta\nu}&=\int\frac{d^4q}{(2\pi)^4}g^{\mu\sigma}[g_{D^\ast D^\ast V}(q-p_2)^\beta g^{\eta\epsilon}\nonumber\\
        &+4f_{D^\ast D^\ast V}(p_{4}^{\epsilon}g^{\beta\eta}-p_{4}^{\eta}g^{\beta\epsilon})]\nonumber\\
        &\times[2ie\epsilon^{\alpha\nu\delta\lambda}v_{\delta}(\frac{Q\beta^{\prime}}{2}+\frac{Q^{\prime}}{2m_c})p_{3\alpha}]S(p_1, m_D)\nonumber\\
        &\times S_{\sigma \eta}(p_2, m_{D^\ast})S_{\lambda\epsilon}(q,m_{D^\ast})F(q^2,m_{D}^{2}),
    \end{align}
    \begin{align}
        I_{(c)}^{\mu\beta\nu}&=\int\frac{d^4q}{(2\pi)^4}g^{\mu\omega}[-g_{DDV}(q-p_2)^\beta]\nonumber\\
        &\times[2iev_{\delta}(\frac{Q\beta^{\prime}}{2}+\frac{Q^{\prime}}{2m_c})\epsilon^{\alpha\nu\delta\tau}p_{3\alpha}] S_{\omega\tau}(p_1, m_{D^\ast})\nonumber\\
        &\times S(p_2, m_{D})S(q,m_D)F(q^2,m_{D}^{2}),
    \end{align}
    \begin{align}
        I_{(d)}^{\mu\beta\nu}&=\int\frac{d^4q}{(2\pi)^4}g^{\mu\omega}[-2f_{D^\ast DV}\epsilon^{\alpha\beta\delta\epsilon} p_{4\alpha}(q-p_2)_{\delta}]\nonumber\\
        &\times eQ_{D^\ast}[(p_1+q)^\nu g^{\tau\lambda}-q^{\tau}g^{\lambda\nu}-p_1^{\lambda}g^{\tau\nu}]S(p_2, m_{D})\nonumber\\
        &\times S_{\omega\tau}(p_1, m_{D^\ast})S_{\lambda\epsilon}(q,m_{D^\ast})F(q^2,m_{D}^{2}),
    \end{align}
    \begin{align}
        I_{(e)}^{\mu\beta\nu}&=\int\frac{d^4q}{(2\pi)^4}g^{\mu\sigma}[- g_{DDV}(p_1 +q)^\beta]\nonumber\\
        &\times [2iev_{\delta}(\frac{Q\beta^{\prime}}{2}+\frac{Q^{\prime}}{2m_c})\epsilon^{\alpha\nu\delta\zeta}p_{3\alpha}]S(p_1, m_D)\nonumber\\
        &\times S_{\sigma \zeta}(p_2, m_{D^\ast})S(q,m_D)F(q^2,m_{D}^{2}),
    \end{align}
    \begin{align}
        I_{(f)}^{\mu\beta\nu}&=\int\frac{d^4q}{(2\pi)^4}g^{\mu\sigma}[2f_{D^\ast DV}\epsilon^{\alpha\beta\delta\epsilon} p_{4\alpha}(p_1 +q)_{\delta}]\nonumber\\
        &\times e Q_{D^\ast}[(q-p_2)^\nu g^{\lambda\zeta}+p_{2}^{\lambda}g^{\zeta\nu}-q^{\zeta}g^{\lambda\nu}]S(p_1, m_D)\nonumber\\
        &\times S_{\sigma \zeta}(p_2, m_{D^\ast})S_{\lambda\epsilon}(q,m_{D^\ast})F(q^2,m_{D}^{2}),
    \end{align}
    \begin{align}
        I_{(g)}^{\mu\beta\nu}&=\int\frac{d^4q}{(2\pi)^4}g^{\mu\omega}[eQ_D(p_2 -q)^{\nu}]\nonumber\\
        &\times[-2f_{D^\ast DV}\epsilon^{\alpha\beta\delta\tau} p_{4\alpha}(p_1 +q)_{\delta}]S_{\omega\tau}(p_1, m_{D^\ast})\nonumber\\
        &\times S(p_2, m_{D})S(q,m_D)F(q^2,m_{D}^{2}),
    \end{align}
    \begin{align}
        I_{(h)}^{\mu\beta\nu}&=\int\frac{d^4q}{(2\pi)^4}g^{\mu\omega}[2iev_{\delta}(\frac{Q\beta^{\prime}}{2}+\frac{Q^{\prime}}{2m_c})\epsilon^{\alpha\nu\delta\lambda}p_{3\alpha}]\nonumber\\
        &\times[g_{D^\ast D^\ast V}(p_1 +q)^\beta g^{\epsilon\tau}+4f_{D^\ast D^\ast V}(p_{4}^{\tau}g^{\beta\epsilon}-p_{4}^{\epsilon}g^{\beta\tau})]\nonumber\\
        &\times S_{\omega\tau}(p_1, m_{D^\ast})S(p_2, m_{D})S_{\lambda\epsilon}(q,m_{D^\ast})F(q^2,m_{D}^{2}).
    \end{align}
    
Here the propagators for the charmed mesons $D$ and $D^\ast$ are
    \begin{align}
        S(q,m_D) &=\frac{1}{q^2-m_{D}^{2}+i\epsilon}\,,\\
        S^{\mu \nu }(q,m_{D^\ast}) &=\frac{-g^{\mu\nu}+q^\mu q^\nu /m_{D^\ast}^2}{q^2-m_{D^\ast}^2 +i\epsilon}.
    \end{align}

The radiative decay width of the $X(3872) \rightarrow \gamma V$ is given by 
\begin{align}
\Gamma(X(3872)\rightarrow \gamma V)=\frac{E_{\gamma}{\left\vert \mathcal{M}_{X} \right\vert}^2}{24\pi M_{X}^2},
\end{align}
where $E_\gamma$ is the photon energy in the $X(3872)$ rest frame.

\section{NUMERICAL RESULTS}\label{sec:numerical results}

Before we present the numerical results for the radiative decay of $X(3872)$, the relevant coupling constants and form factors will be addressed firstly. The coupling constants in Eq. (\ref{eq:LX(3872)}) were determined as
\begin{align}\label{gnc}
        &g_n=\left\vert g_{\mathrm{eff}}^{n}\right\vert \cos\theta,\nonumber\\
        &g_c=\left\vert g_{\mathrm{eff}}^{c}\right\vert \sin\theta,
\end{align}
where the $g_{\mathrm{eff}}^{n/c}$ are given by~\cite{Albaladejo:2022sux,Baru:2003qq}
\begin{align}
        g_{\mathrm{eff}}^{2}=16\pi M (m_1+m_2)\sqrt{\frac{2\left\vert M-m_1-m_2\right\vert}{m_1m_2/(m_1+m_2)}}. 
\end{align}

Using the masses of the $X(3872)$ and charmed mesons, we can obtain the value of $g_{\mathrm{eff}}$ as follows
\begin{align}\label{geff}
        &\left\vert g_{\mathrm{eff}}^{n}\right\vert =2.77~ \mathrm{GeV},\nonumber\\
        &\left\vert g_{\mathrm{eff}}^{c}\right\vert =9.94~ \mathrm{GeV}.
\end{align}

It is generally accepted that the form factor can be used to describe the inner structures and off-shell effects of the charmed mesons in the hadronic loops. It is recalled that the mass of $X(3872)$ is near the thresholds of $\bar{D}^\ast D$. Consequently, it makes clear that the components of $X(3872)$, i.e.
the charmed mesons in the loop connected to $X(3872)$, are almost on-shell. 

For the radiative decay processes studied in this work, we adopt dipole form factor to depict the inner structure and the off-shell effects of the exchanged charmed meson, which is \cite{Li:1996yn,Cheng:2004ru}
\begin{align}\label{FF}
F(q^2,m^2)=\left(\frac{m^2-\Lambda^2}{q^2-\Lambda^2} \right)^2,
\end{align}
where $q$ and $m$ are the momentum and mass of the exchanged meson, respectively \cite{Cai:2024glz}. The parameter $\Lambda$ can be further reparameterized as $\Lambda=m+\alpha\Lambda_{QCD}$ with $\Lambda_{QCD}=0.22~\mathrm{GeV}$. The model parameter $\alpha$ should be of order of unity \cite{Wang:2022qxe,Wu:2021udi,Zheng:2024eia}. However, its concrete value cannot be estimated from first principle. In general, the value of $\alpha$ is determined by comparing theoretical estimates with experimental measurements. Its value depends not only on the exchanged particles but also on the external particles involved in the strong interaction \cite{Zheng:2024eia}. In the present calculations, we vary the value of $\alpha$ from $0.5$ to $2.0$ to illustrate the model dependence of the decay processes under consideration.

\begin{figure*}[htbp]
    \begin{center}
    \includegraphics[width=3.5in]{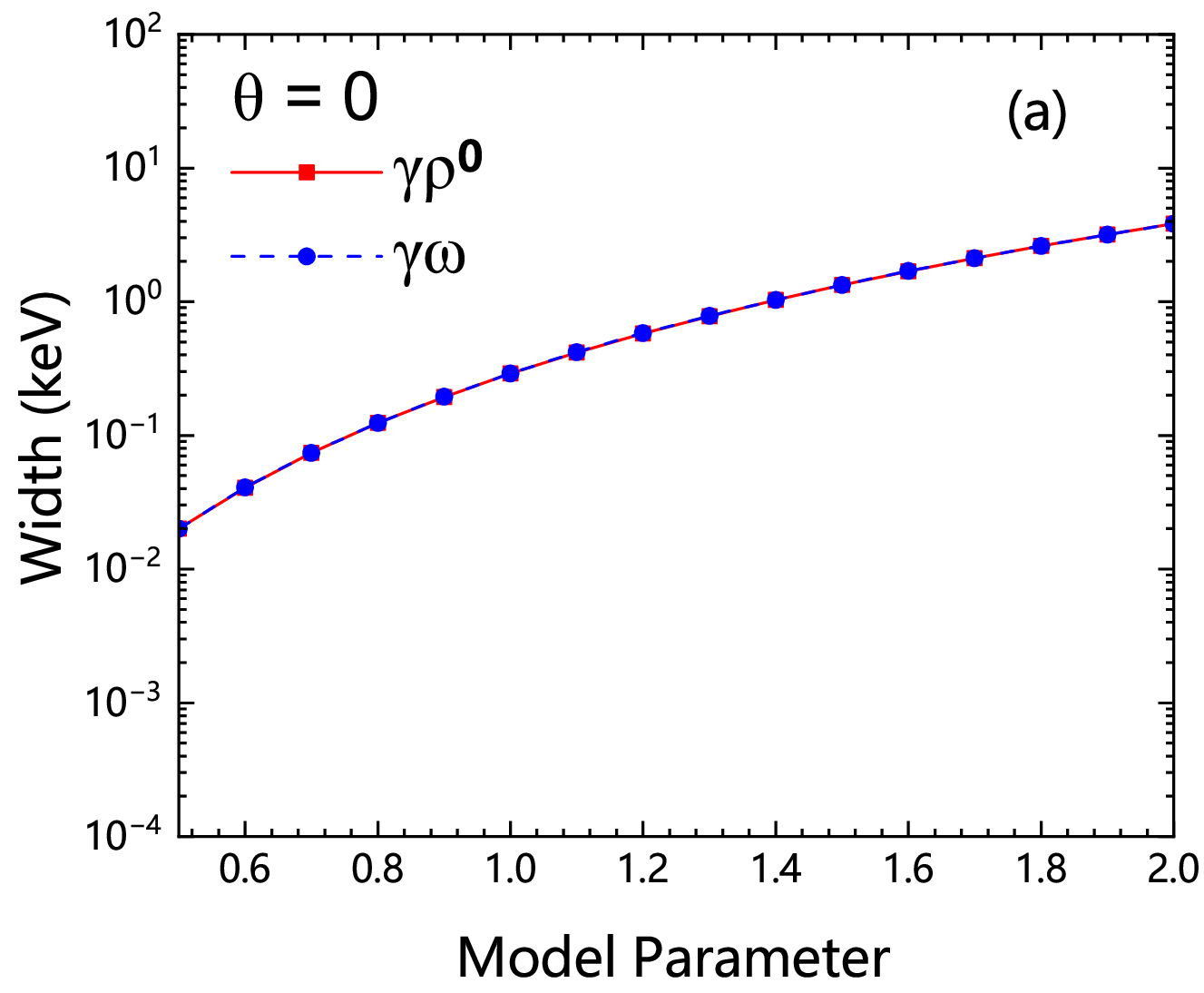}
    \includegraphics[width=3.5in]{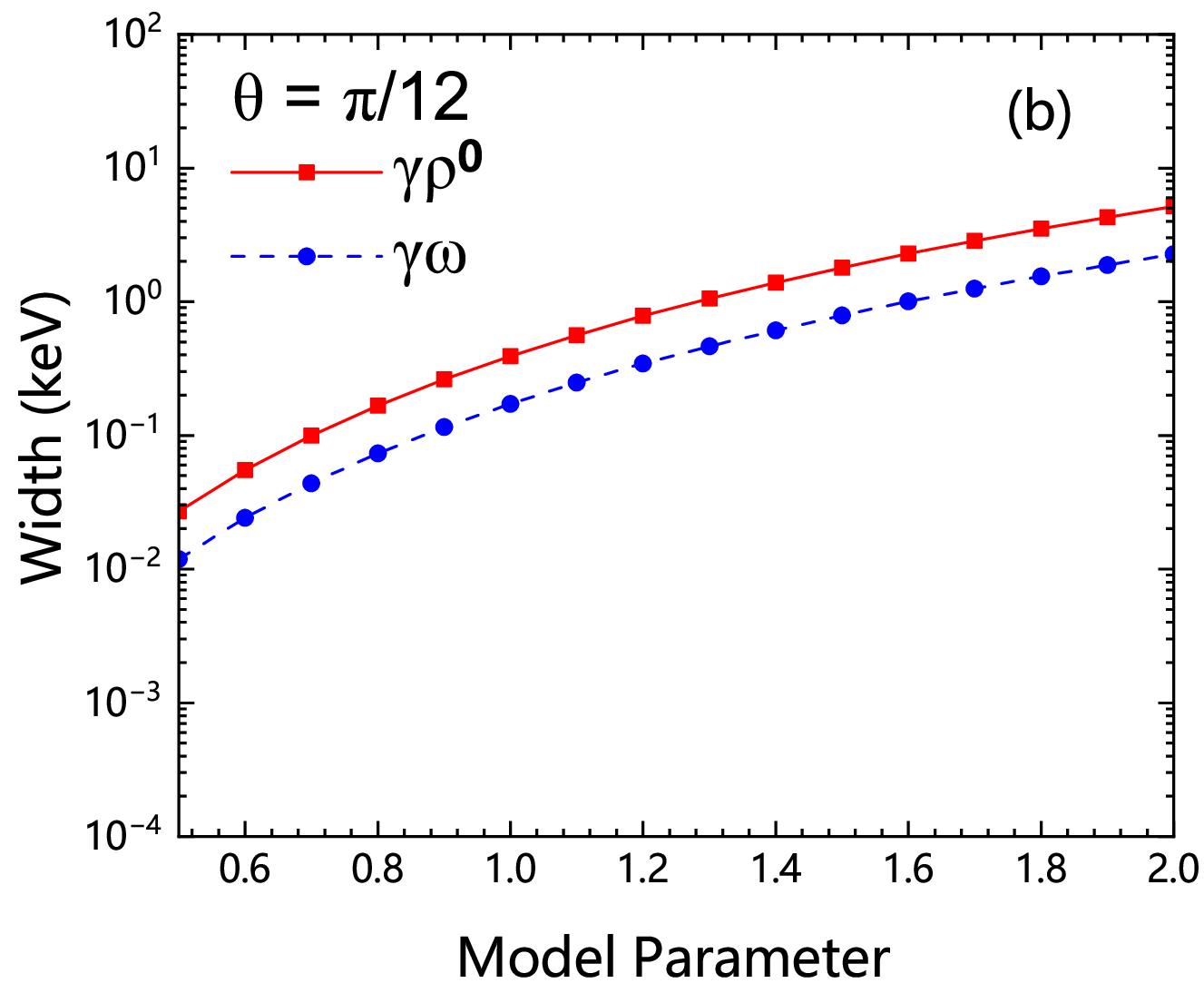}\\
    \includegraphics[width=3.5in]{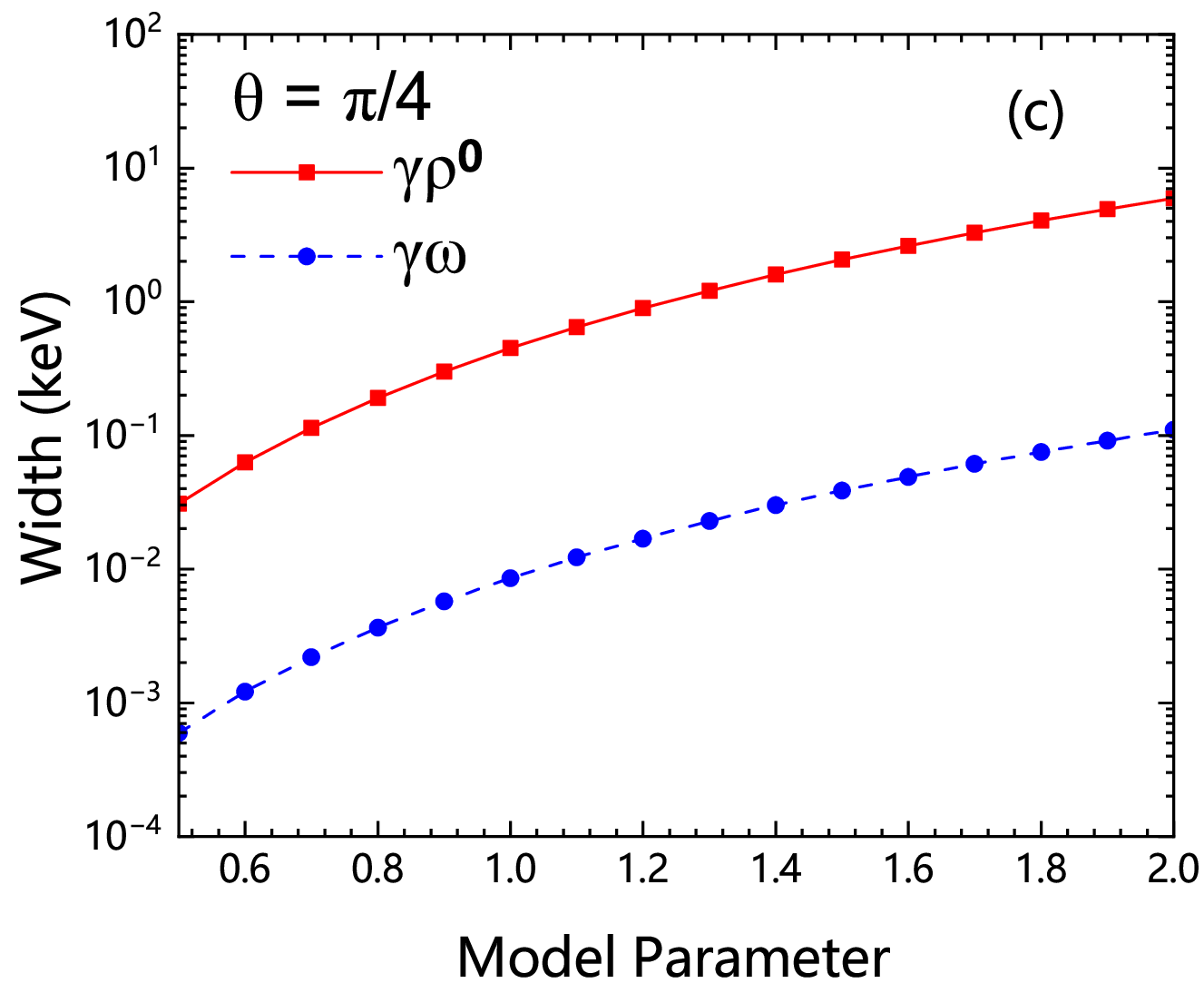} 
    \includegraphics[width=3.5in]{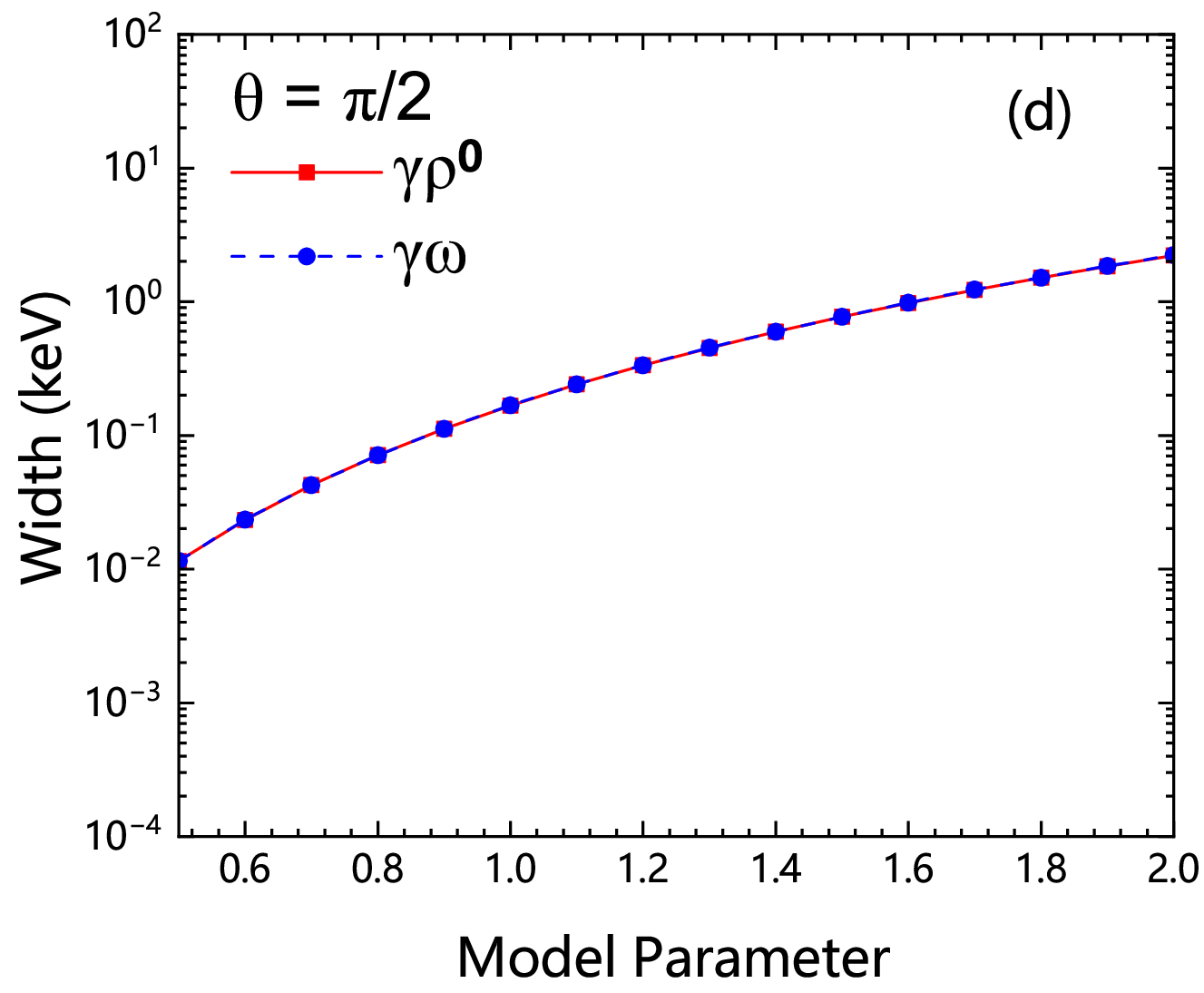}
    \end{center}
    \caption{Widths of the $X(3872) \rightarrow \gamma \rho^0$ and  $X(3872) \rightarrow \gamma \omega$ as a function of the model parameter $\alpha$. The $\theta$ is taken to be $0$, $\pi/12$, $\pi/4$, and $\pi/2$, respectively. }
    \label{fig:anglerange}
\end{figure*}

\begin{table*}
	\caption{The decay widths (in units of keV) and the ratio $R$ defined in Eq.~(\ref{Rratio}) for the $X(3872) \rightarrow \gamma V ~(V=\rho^0 ,\, \omega)$ with different $\theta$ values. The $\alpha$ range is $0.5\sim2.0$.}
	\label{tab:widths}
	\begin{ruledtabular}
		\begin{tabular}{cccc}
			Phase angle	&$\Gamma[X(3872) \rightarrow \gamma \rho^0]$  &$\Gamma[X(3872) \rightarrow \gamma \omega]$ & R\\
			\colrule
			$0$ & $0.02\sim3.83$ &  $0.02\sim3.84$ & $\sim1$\\
			$\pi/12$ & $0.03\sim5.18$ & $0.01\sim2.27$ & $\sim2.3$\\
			$\pi/4$ & $0.03\sim5.94$ &$0.001\sim0.11$ & $\sim53$\\
			$\pi/2$ & $0.01\sim2.22$ &$0.01\sim2.23$ & $\sim1$\\
		\end{tabular}
	\end{ruledtabular}
\end{table*}

In the following, we investigate the radiative decay widths of $X(3872) \rightarrow \gamma \rho^0$ and  $X(3872) \rightarrow \gamma \omega$ in terms of the model parameter introduced in Eq.~(\ref{FF}), which serves as the only model parameter in our calculation. In Fig.~\ref{fig:anglerange}, we plot the radiative decay widths of $X(3872) \rightarrow \gamma \rho^0$ and  $X(3872) \rightarrow \gamma \omega$ with $\alpha$ varying from $0.5$ to $2.0$ To examine the influence of the neutral and charged constituents in $X(3872)$ on the radiative decay widths, we choose four typical values of the phase angle: $\theta=0,\pi/12,\pi/4$, and $\pi/2$. 

In the case of $\theta =0$, the $X(3872)$ is a pure bound state with only neutral components. On the contrary, for $\theta =\pi/2$, the $X(3872)$ is a pure bound state with only charged components. As shown in Figs. \ref{fig:anglerange} (a) and (d), the radiative decay width of the $X(3872) \rightarrow \gamma \rho^0$ is nearly equal to that for the $X(3872) \rightarrow \gamma \omega$. In these two cases, the radiative decay widths are on the order of $0.01\sim5$ keV.

When $\theta =\pi/12$, the isospin violation is minimal for the radiative decay of the $X(3872)$. Both neutral and charged components exist at $\theta =\pi/12$, but the neutral component dominates. In this case, the couplings of the $X(3872)$ to the neutral and charged components are nearly equal, i.e., $g_n\approx g_c$. It is seen from Fig. \ref{fig:anglerange} (b) that the radiative decay width of $X(3872) \rightarrow \gamma \rho^0$ is about three times larger than that for $X(3872) \rightarrow \gamma \omega$. 
For $\theta =\pi/4$, the neutral and charged components in the $X(3872)$ have equal proportions. As shown in Fig. ~\ref{fig:anglerange}(c), the radiative decay widths of the $X(3872) \rightarrow \gamma \rho^0$ is about fifty times larger than that for $X(3872) \rightarrow \gamma \omega$.

From Fig.~\ref{fig:anglerange}, we can see that the decay widths for both $X(3872) \rightarrow \gamma \rho^0$ and $X(3872) \rightarrow \gamma \omega$ processes exhibit similar $\alpha$-dependence, showing a  monotonic increase with increasing parameter $\alpha$. The width ratio between these two channels can effectively mitigate or even cancel out the model dependence, making it a robust observable \cite{Wang:2022qxe}. However, Fig.~\ref{fig:anglerange} shows that the trend of these two processes with the phase angle are opposite generally. The ratio of decay widths provides a clear demonstration that how the phase angle $\theta$ affects the decay dynamics. Specifically, the width ratio between the two different radiative decays can help us to further discuss the uncertainties arising from the introduction of form factors \cite{Cai:2024glz,Zheng:2024eia,Wang:2022qxe}. For the radiative decay $X(3872) \rightarrow \gamma V$, we define the following width ratio $R$:
\begin{align}
    R=\frac{\Gamma (X(3872) \rightarrow \gamma \rho^0)}{\Gamma (X(3872) \rightarrow \gamma \omega)}. \label{Rratio}
\end{align}

\begin{figure}
    \centering
    \includegraphics[width=1.0\linewidth]{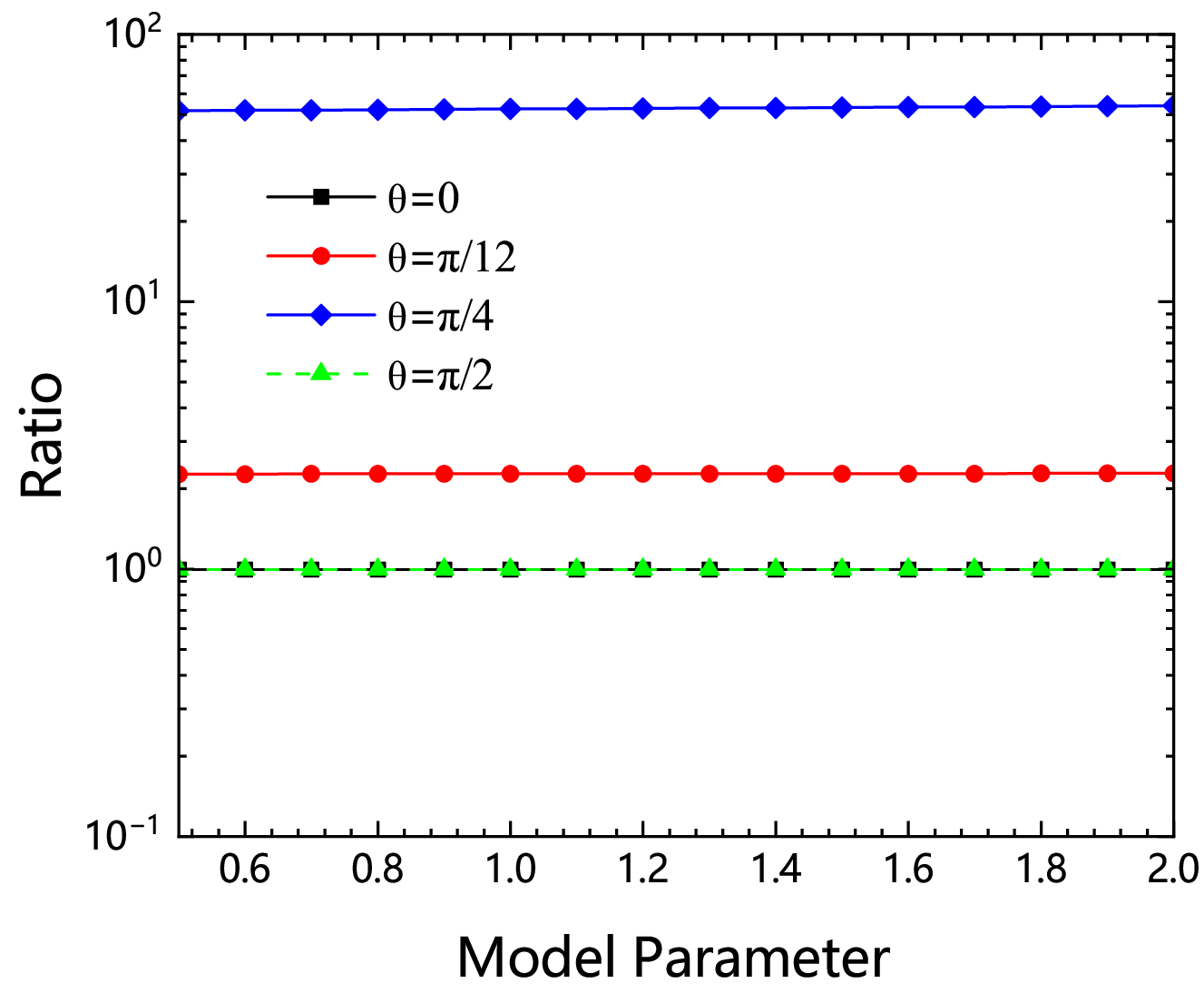}
    \caption{The model parameter $\alpha$ dependence of the ratio $R$ defined in Eq.~(\ref{Rratio}).}
    \label{fig:ratio}
\end{figure}

In Fig.~\ref{fig:ratio}, we present the width ratio $R$ as a function of the model parameter $\alpha$. The results clearly show that the width ratios are 
rather insensitive to the model parameter $\alpha$. On the contrary, there is a strong dependence of the width ratio $R$ on the phase angle $\theta$, which reflects the sensitivity to the $R$ to the inner structure of the $X(3872)$, characterized by the proportion of the neutral and charged molecular components in $X(3872)$. We hope that the decay width of $X(3872) \rightarrow \gamma V ~(V=\rho^0 , \omega)$, especially their ratio, could be measured in the future, which is a crucial test of the molecular nature of $X(3872)$. The values of the decay widths and their ratios for $X(3872) \rightarrow \gamma V ~(V=\rho^0 , \omega)$ at four $\theta =0,\,\pi/12,\, \pi/4$, and $\pi/2$ are shown in Tab.~\ref{tab:widths} with $\alpha$ ranging from $0.5$ to $2.0$.

\begin{figure}
    \centering
    \includegraphics[width=1.0\linewidth]{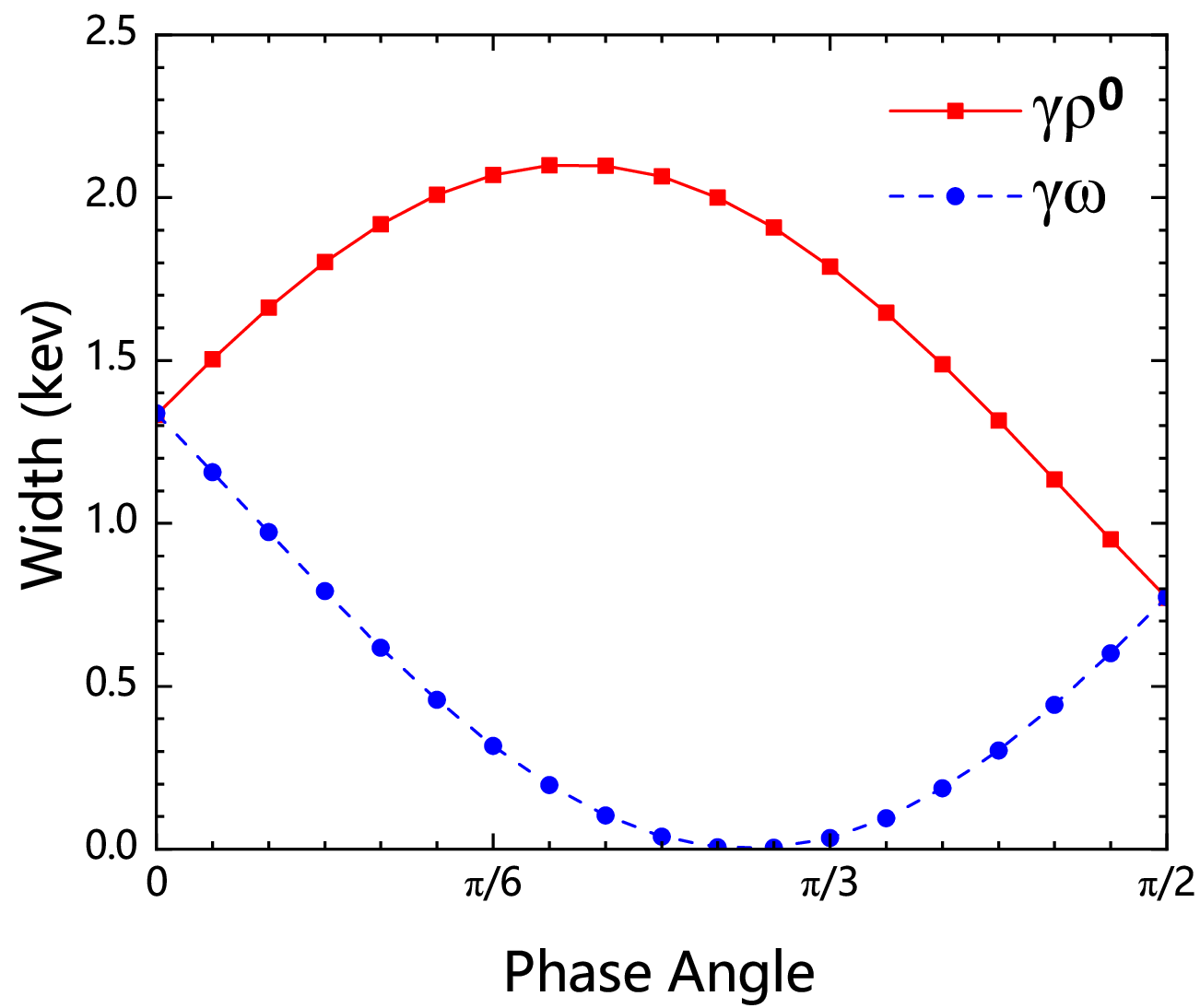}
    \caption{Widths of the $X(3872) \rightarrow \gamma \rho^0$ and  $X(3872) \rightarrow \gamma \omega$ as function of the phase angle. The model parameter $\alpha$ is taken to be $1.5$.}
    \label{fig:width_angle}
\end{figure}

\begin{figure}
    \centering
    \includegraphics[width=1.0\linewidth]{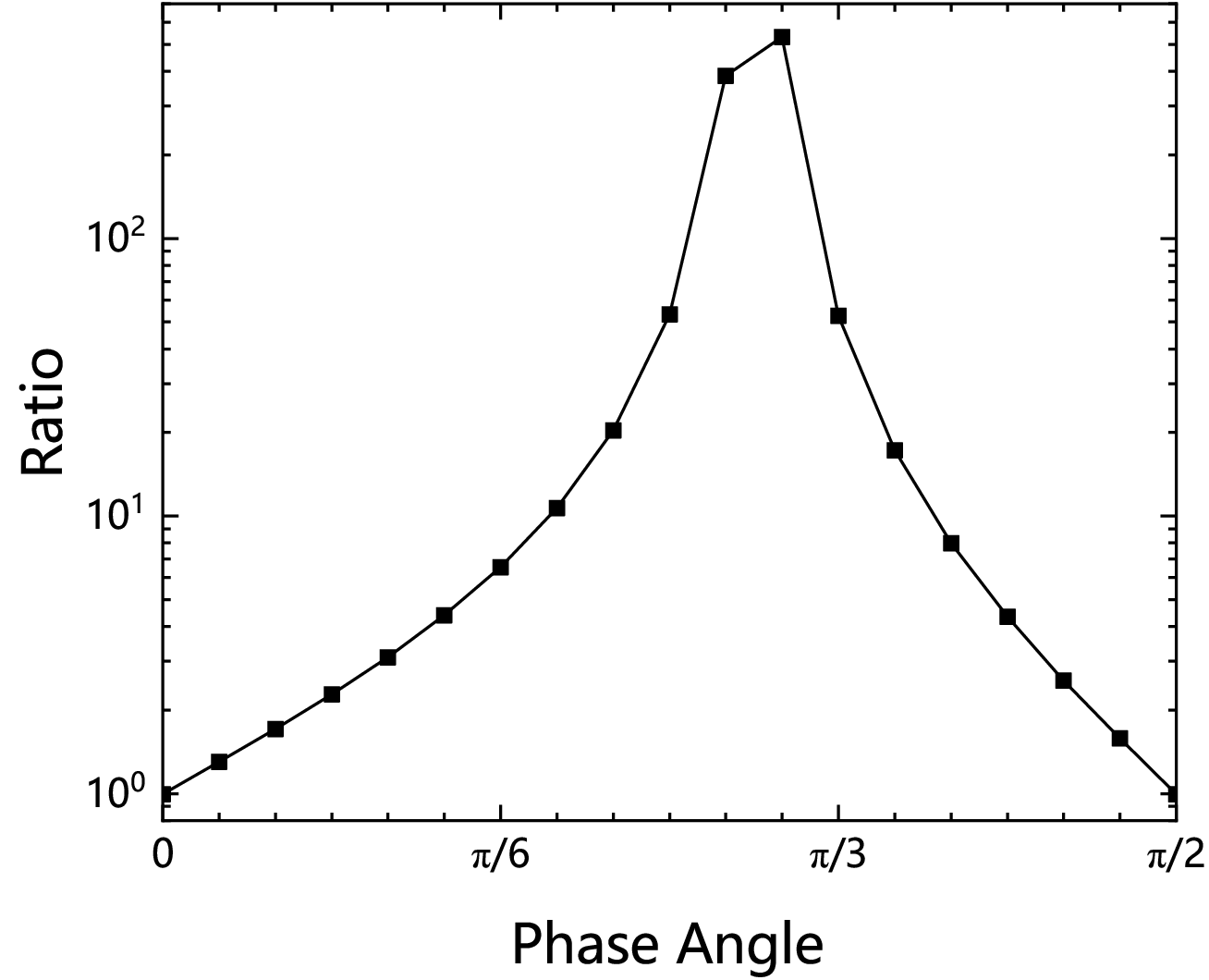}
    \caption{The phase angle $\theta$ dependence of the ratio $R$ defined in Eq.~(\ref{Rratio}). The model parameter $\alpha$ is taken to be $1.5$.}
    \label{fig:width_ratio}
\end{figure}

To clearly show the influence of the phase angle $\theta$, we plot in Figs.~\ref{fig:width_angle} and \ref{fig:width_ratio} the dependence of the decay width and their ratio on the phase angle for $\alpha=1.5$. For $X(3872) \rightarrow \gamma \rho^0$, the decay width first increases and then decreases as a function of the phase angle, while the behavior is opposite for $X(3872) \rightarrow \gamma \omega$. From Fig.~\ref{fig:width_angle}, one can see that the decay widths of $X(3872) \rightarrow \gamma \rho^0$ and $X(3872) \rightarrow \gamma \omega^0$ reach their maximum and minimum, respectively, around $\theta=\pi/4$. As a result, the ratio first increases and then decreases with the increase of phase angle, as shown in Fig.~\ref{fig:width_ratio}. The ratio shown in Fig.~\ref{fig:width_ratio} may be tested by the future experimental measurements and can be used to determine the value of the phase angle.

\section{SUMMARY}\label{sec:summary}

In this work, we investigate the radiative decays of the $X(3872)$ to light hadrons in the molecule scenario, where the $X(3872)$ is regarded as a pure hadronic molecule of the $D\bar{D}^*+c.c$ in an $S$-wave with the quantum numbers $J^{PC}=1^{++}$. For the molecule state $X(3872)$, we considered four cases, i.e., pure neutral components ($\theta=0$), minimal isospin breaking effects ($\theta=\pi/12$), isospin singlet ($\theta=\pi/4$) and pure charged components ($\theta=\pi/2$), where $\theta$ is a phase angle describing the proportion of neutral and charged constituents. When the $X(3872)$ is a purely neutral or charged $D{\bar D}^*+c.c$ bound
state, the predicted partial decay widths of $X(3872)\to \gamma \rho^0$ and $\gamma \omega$ can reach about several keV, and the corresponding branch ratio is about $10^{-2}$. When there are both neutral and charged components in $X(3872)$, the predicted partial decay widths of $X(3872)\to \gamma \rho^0$ can reach a few tens of keV, while it is several keV for $X(3872)\to \gamma \omega^0$.  
 
 It is found that the absolute decay widths are model-dependent, but the relative width ratio is rather independent of the model parameter. Moreover, the calculated results indicate that the radiative decays of the $X(3872)$ are strongly influenced by the molecular configuration characterized by the proportion of the charged and neutral constituents. We hope that the present calculations could be tested by the experimental measurements.		

The radiative decay of $X(3872)$ is a key indicator of its physical properties and internal structure. By studying its radiative decay, we can further explore the internal structure and properties of the $X(3872)$. The study of the $X(3872)$ not only reveals new phenomena in hadron physics, but also provides a new perspective for our in-depth understanding of hadronic matter.

\begin{acknowledgments}\label{sec:acknowledgements}
This work is partly supported by the National Natural Science Foundation of China under Grant Nos. 12475081, 12105153, 12405093, and by the Natural Science Foundation of Shandong Province under Grant Nos. ZR2021MA082 and ZR2022ZD26. It is also supported by Taishan Scholar Project of Shandong Province (Grant No.tsqn202103062) and by National Key Research and Development Program under Grant No. 2024YFA1610504.	
\end{acknowledgments}

\twocolumngrid
\bibliography{reference.bib}
\end{document}